# Medium dependence
# of asphaltene agglomeration inhibitor efficiency

**Mariana Barcenas, Pedro Orea, Eduardo Buenrostro-González,
Luis S. Zamudio-Rivera, and Yurko Duda***

Programa de Ingeniería Molecular, Instituto Mexicano del Petróleo,
Eje Central L.Cardenas 152, 07730, México D.F., México;
*E-mail: yduda@lycos.com

Applying chemical additives (molecule inhibitors or dispersants) is one of the common ways to control asphaltene agglomeration and precipitation. However, it is not clear why at some conditions the synthetic flocculation inhibitors as well as resins not only do not inhibit the asphaltene agglomeration, they may also promote it, and why the increasing of the additive concentration may lead to the diminishing of their efficacy. To clarify this issue, in the present work we have performed a set of vapor preassure osmometry experiments investigating the asphaltene agglomeration inhibition by commercial and new inhibitor molecules in toluene and o-diclorobenzene. Monte Carlo computer modeling has been applied to interpret some unexpected trends of molar mass of the Puerto Ceiba asphaltene clusters at different concentrations of inhibitor, assuming that inhibitors efficiency is directly related to their adsorption on the surface of asphaltene or its complexes. It has been found that a self-assembly of inhibitor molecules, induced by relative lyophilic or lyophobic interactions, may be a reason of the inhibitor efficacy declining.

## Introduction

Asphaltenes are the most intriguing fraction of petroleum fluids. Polarity and complex structure of asphaltenes determine the viscosity of crude oils as well as their tendency to flocculate and precipitate during the course of oil recovery and refinery. Asphaltenes comprise the heaviest and most polar fraction of crude oils and are broadly classified as the solubility class of components that are insoluble in low molecular weight alkanes, such as n-hexane. It is widely held that asphaltenes exist in the form of colloidal dispersions and are stabilized in solution by resins and aromatics, which act as peptizing agents [1-7]. Resins are believed to stabilize asphaltenes by bridging between polar asphaltene particles and the nonpolar oil surrounding them. The polar sites of the resins interact with polar asphaltene sites, while the nonpolar (alkyl) sites of the resins interact with the bulk oil phase. According to some authors the asphaltenes precipitation by an aliphatic solvent is related to the resins desorption on the surface of the asphaltenes colloidal particles [6-11] or dictated by changes in the solubility parameter of crude oil [12,13] which is influenced by pressure, temperature and composition changes during oilfield operations.

The deposition and precipitation of flocculated asphaltenes can severely reduce the permeability of the reservoir, and also plug-up the well bore and tubing [14]. Therefore, the development of a cost-effective method to control the asphaltene flocculation and deposition (various production and chemical treatment techniques) is of great importance to increase the overall efficiency of the oil recovery in the fields with asphaltene problems.

Since an important fact about asphaltenes is they are deposited only after flocculation, applying of synthetic inhibitors to prevent asphaltene flocculation has been studied extensively [15-22]. These molecules have characteristics in common with petroleum resins and are believed to interact with asphaltenes and oil in a similar manner. It is well accepted that the inhibitors are surface active, in the sense that they adsorb to the asphaltene surface or small asphaltene clusters, and thereby disrupt further aggregation of them. Leon *et al*. [4] stated that the activity of the asphaltene flocculation inhibitors (AFI) is related to the maximum amount of AFI adsorbed on the asphaltene surface. According to this, the larger the AFI concentration at the asphaltene surface, the larger the volume of n-heptane needed to begin the flocculation of asphaltenes [4].

It was shown [15], that the activity of inhibitors is not only dependent on the acidic head, but also of the aliphatic or aromatic tail of the AFI molecule. Although, recent experimental works have revealed the features that an efficient inhibitor needs to possess [15-22], the overall physical mechanism that governs the inhibition is far from being understood. For example, it is not clear why in some cases the resins and dispersants not only do not inhibit precipitation, they may also promote it, and why the increasing of AFI's concentration may lead to diminishing of their efficacy.

To clarify these issues, in the present work we have performed a set of experiments investigating the asphaltene agglomeration inhibition by commercial and two new inhibitor molecules [22] in toluene and o-diclorobenzene (ODCB). Molecular modeling has been applied to interpret qualitatively some unexpected trends of the mean cluster size of asphaltene aggregates at different concentration of inhibitors.

## Materials and Experimental Methods

**Materials.** The asphaltene sample used in this study has been extracted from a crude oil labeled as Puerto Ceiba (PC) from the southern production region of Mexico at the Tabasco state. This crude oil is a highly unstable one with relatively low asphaltene content, which present strong flocculation-deposition problems during production process. Some characteristics of this oil are given in Table 1.

The asphaltenes were isolated from the crude oil by addition of *n*-heptane in a ratio of 40:1 (cm$^3$ per g). The suspension was left under strong mixing by 8 h, then filtered using a 1 μm porous size membrane and dried under vacuum at 60$^o$C overnight. In order to remove *n*-heptane soluble material trapped in the solids, they were re-dissolved in methylene chloride (10 wt %) and re-precipitated with *n*-heptane using the same precipitant-solution ratio of 40:1. The insoluble material was separated by filtration. Fresh *n*-heptane was added to the solids in order to prepare a suspension. The suspension was put under 20-min ultrasonic shaking followed by a centrifugation process, which separates the supernatant (i.e. *n*-heptane + *n*-heptane solubles) from the *n*-heptane insolubles. The supernatant was separated by decantation and fresh *n*-heptane was added again to prepare a new suspension. This shaking-centrifugation cycle was repeated until the supernatant became transparent with a pale yellow color that did not change after various washing cycles.

**Amphiphile molecules.** The commercial inhibitor was the 4-nonylphenol (NP), from Aldrich analytical grade, purity ⩾98% used without further purification and two new inhibitor molecules DAIM-2000 (hereafter denoted as M2) and DAIM-3000 (hereafter as M3) are oxazolines derivative of poly alkyl or poly alkenyl N-hydroxyalkyl, which are reported in [22].

**Vapor Pressure Osmometry (VPO)**. The principle of vapor pressure osmometry is based on the difference in vapor pressure caused by the addition of a small amount of solute to a pure solvent. In the vapor pressure osmometer, a droplet of pure solvent and a droplet of solvent-solute are placed on separate thermistors surrounded by pure solvent vapor. The difference in vapor pressure between the two droplets results in a difference in temperature at each thermistor. The temperature difference causes a voltage difference, which is related to the molar mass (MM), $m_2$, of the solute as follows,

$$\Delta V = \frac{K \cdot C_2}{m_2},$$

where $\Delta V$ is the voltage difference between the thermistors, $C_2$ is the concentration of the solute, K is the calibration constant. To calibrate the instrument, solutes that form nearly ideal mixtures with the solvent at low concentrations are chosen. As the solute MM is known, the calibration constant can be determined from the intercept of a plot of $\Delta V/C_2$ versus $C_2$.

Once the instrument is calibrated, the voltage differences for solutions of asphaltenes and pure solvent can be measured. The asphaltene MM can be found from the intercept of a plot of $\Delta V/C_A$ versus $C_A$, where $C_A$ is the asphaltene concentration [23-25].

The asphaltene MM in toluene and toluene-inhibitor systems was measured with a Model Wescan 232 VPO. This osmometer has a detection limit of $5 \times 10^{-6}$ mol/L when used with toluene. Benzil was used to calibrate the instrument and a polymer standard was used to check the calibration. Ten to fifteen readings were taken at each concentration to obtain the voltage response for that concentration.

## Theoretical Modeling

In this work we propose a new simple model of asphaltene flocculation inhibition. The model is studied by means of canonical Monte Carlo (MC) simulation method [26]. The simulations are performed on a two-dimensional (2D) plane for reasons of computational efficiency and to facilitate the visual identification of the peculiar configurations [27,28]; in spite of such simplification, we are confident that our conclusions are qualitatively valid for the three-dimensional counterpart.

We consider the surfaces of two asphaltene colloids as hard impenetrable walls separated by distance $L_X$. Each of the walls possesses $n_a$ associative sites, which mimic chemically active sites of asphaltene colloid surface [12]. These two colloids are immersed in a fluid mixture of $N_W$ solvent and $N_I$ inhibitor molecules, see fig.1 for details.

The inhibitor molecule is modeled as a chain that consists of a head (H) disc and m interconnected hard discs of hydrocarbon tail (T) group, with diameters $\sigma_H$ and $\sigma_T$, respectively. These molecules are dissolved in a solvent, which is modeled by hard discs of diameter $\sigma_W$, that is chosen to be a unit length, $\sigma_W = \sigma_T = 1$; the H-part diameter is $\sigma_H = 1.2$.

The particle-particle pair interaction of the inhibitor parts with the solvent is given by

$$\beta U_{iW}(r) = \begin{cases} \infty, & \text{if } r < \sigma_{iW} \\ 0, & \text{if } r > \sigma_{iW} \end{cases}, \quad (1)$$

where the subscript $i$ stands for H or T; $r$ measures the particles separation, and $\sigma_{iW} = 0.5(\sigma_i + \sigma_W)(1+\Delta_i)$, where $\Delta_i$ is the non-additivity parameter. A positive non-additivity parameter, $\Delta_i > 0$, in the cross interaction between two species, H-W or T-W, introduces some kind of an 'additional repulsion' between unlike particles. A variation of these parameters is equivalent to changing the solubility of the AFI in the specific solvent. In other words, positive $\Delta_H$ and $\Delta_T$ parameters self-consistently account for the lyophobicity of AFI head and tail, respectively. That is, the higher value of $\Delta_i$, the more lyophobic i-W interaction, and as a consequence, the lower solubility of the particle i [27-29].

In order to mimic varying flexibility of tail chains, in previous works [27] we applied the so-called fused hard-sphere chain model. In the present work, for this purpose, we consider nonadditive interaction between no adjacent elements of the same chain: That is, the distance between them is limited by value of R, $R = \sigma_T + \Delta_{TT}$, where $\Delta_{TT}$ is the factor which controls the chain flexibility. By changing the parameter $\Delta_{TT}$, one can deal with chains of different degree of flexibility, i.e. from totally flexible chains, $\Delta_{TT} = 0$, up to completely rigid chains $\Delta_{TT} = \sigma_T$ [28]. In the present work we use $\Delta_{TT} = 0.4$ and $0.8$. Changing of this parameter does not alter qualitatively our conclusions as will be explained below.

The interaction between particles of each species, H, T, and W, with the walls is of a hard core type in the $x$ direction,

$$\beta U_0(x) = \begin{cases} \infty, & \text{if } L_X < x, \text{ or } x < 0 \\ 0, & \text{if } L_X > x > 0 \end{cases}, \quad (2)$$

where $x$ is the distance between the centre of each particle and the surface in the direction perpendicular to the surface.

In order to mimic the adsorption of AFI molecules on the asphaltene surface, the H-part of the inhibitor molecules associates with an attractive site due to the following pair potential [28,30],

$$\beta U_{as}(r) = \begin{cases} -\varepsilon, & \text{if } \lambda > r, \\ 0, & \text{if } \lambda < r \end{cases}, \qquad (3)$$

Thus, the associative attraction which acts between the asphaltene surface site and the center of the H-particle, is explicitly given by the square-well potential. The appropriate choice of its range, $\lambda = 0.2$ (for the present calculations), guarantees that the saturation condition is satisfied, i.e. only one H-particle can be bonded to each asphaltene surface site [10]. Potential well depth, $\varepsilon$, is chosen to be 4 in this study.

In order to qualify the efficiency of inhibition of AFI molecules we have calculated, under equilibrium, the averaged value of the number of occupied surface sites, $<n_a^H>$. Thus, we assume that there is a direct relationship between AFI adsorption and asphaltene dispersion, i.e. the highest ratio, $S = n_a^H/n_a$, indicates the top inhibition efficiency.

Our MC simulations involve from $N_H=30$ to $N_H=120$ chain molecules in the box, that depends on the concentration considered. Unlike an experiment, the AFI number concentration in the simulation is defined as a number of inhibitor molecules per cell area, $X^I = N_H \times (L_X L_Y)^{-1}$ (Note, the inhibitor concentration in our experiments we denote as $X_I$). All the MC calculations have been performed at the same solvent number concentrations, $N_W = 0.39\ L_X L_Y$.

The simulations were started by placing the chains randomly oriented and positioned into the box with no overlapping. Then the system was allowed to evolve, subject to a Metropolis algorithm, until the equilibrium state was reached [26]. The equilibrium state means that the parameters described below are unchanged over sequential $N_H \times 10^8$ MC steps. A MC step consists of the following points: (i) attempts to displace and change the orientation one AFI molecule, (ii) attempts to move one solvent particle.

## Results and Discussion

It is well accepted, that the asphaltene association causes VPO, SANS, laser desorption, and chromatography techniques for measuring molecular weights to give different values of averaged asphaltene molecular weights and sizes, depending on the solvent, temperature, and asphaltene concentration [19,24,31-33]. Since a goal of the present study is to reveal the mechanism of inhibitor performance in different media (and not the defining of the exact asphaltene molecular weight) we apply the VPO technique to make only relative analysis of the averaged molar mass of the PC asphaltene agglomerates.

In Fig. 2 we present our VPO measurements of PC asphaltene MM in toluene at 50°C: As seen, asphaltene MM increases as its concentration augments, reaching a value around MM $\approx 10000$ g/mol at the concentration $C_A \approx 20$ g/L. For comparison, the values of $\sim 3600$ g/mol for B6 crude asphaltene in toluene at 53°C, and $\sim 6000$ g/mol for Athabasca C5-asphaltene in toluene at 50°C have been reported in Refs. [24], and [25], respectively. The PC asphaltene monomer MM defined by the linear extrapolation of the three lowest concentration data, is approximately 975 g/mol.

In fig.2, we also show the influence of two kinds of AFI molecules on the MM of asphaltene clusters in toluene. As viewed, the efficiency of our new inhibitors, M2 and M3, depends strongly on their concentrations. The M2 molecule notably inhibits asphaltene agglomeration at the concentration $X_I = 0.1$ g/L, see fig.2a. It is interesting, that at this concentration, the molecule M2 is efficient for all the asphaltene concentrations considered. Surprisingly, the efficacy of this inhibitor declines if its concentration is augmented. For instance, at $X_I = 0.3$ g/L the inhibition effect is almost negligible.

In the case of the M3 molecule (which is similar to the M2 molecule with only one difference being the more polar head) we observe only small inhibition effect at $X_I = 0.3$ g/L, see fig.2b. Neither increasing nor decreasing of the AFI concentration leads to the reducing of asphaltene aggregates.

To get more insight into the effect of inhibitors on the asphaltene agglomeration, we have performed VPO measurements of the asphaltene MM in toluene and ODCB at different concentrations of the inhibitors. For the sake of convenience, hereafter we will analyse our results in term of the reduced molar mass (RMM), that is, the molar mass of asphaltene clusters in the asphaltene-solvent-inhibitor system divided by their molar mass in the asphaltene-solvent system at each asphaltene concentration considered. In such a way, the pure effect of the inhibitors is highlighted.

In fig. 3, we have shown the values of asphaltene RMM in toluene at T=50$^o$C. These data have been obtained by applying NP and M2 inhibitors, parts (a) and (b) of the figure, respectively. First of all, it is worth noting that small amount of both inhibitors, up to $X_I < 0.2$ g/L, significantly reduces the asphaltene agglomeration. In the case of the inhibitor M2, the efficiency almost does not depends on the asphaltene concentration in the solution, at least for the two values of $C_A$ considered here. However, when we increase the amount of inhibitors, NP or M2, their efficiency suddenly drops down, and RMM value rises. In the case of the NP with $C_A = 6$ g/L we observe some kind of hysteresis, i.e. at the inhibitor concentration $X_I \approx 0.35$ g/L there are two possible values of the RMM, $\approx 0.35$ and $\approx 0.8$. Such unexpected phenomenon could be tentatively explained through the possible metastable states in the system: As will be shown in the theoretical part below, at the same thermodynamic conditions different agglomerate structures can be formed.

To study the possible effect of inhibitor agglomeration on the asphaltene clustering we have also performed VPO measurements of the inhibitor self-association in toluene at T=50$^o$C. As can be seen in fig.4, the MM of both inhibitors, M2 and NP, increases initially up to $X_I = 0.16 \div 0.18$ g/L. After that, the inhibitor MM maintains a constant value, probably indicating formation of associative complexes. Note, that the change of inhibitor MM behavior takes place almost at the same values of $X_I$, at which the inhibition efficacy begins to diminish, fig.3.

More interesting behavior has been found by analysing asphaltene agglomeration in the ODCB solvent at T=90$^o$C. As can bee seen in fig. 5, RMM > 1, which means that both NP and M2 molecules promotes agglomeration of the asphaltenes. As in the previous case of asphaltene-toluene-NP system, here we observe two possible values of the RMM for almost each concentration of the NP molecules. Application of the inhibitor M2 also manifests two possible values of RMM; besides, at $X_I = 0.3$ g/L a drastic augment of the RMM occurs.

VPO measurements performed in the system inhibitor-ODCB at $T=90^oC$ are presented in Fig.6. Here one can observe the similarly unusual behavior of the reduced inhibitor molar mass. First of all, the molar mass of NP consists of two branches, like in fig.5. Second, the MM of the inhibitor M2 manifests strong tendency to rising when $X_I$ increases, maximum value being at $X_I = 0.2$ g/L. It is likely that the two phenomena, the self-association of M2-inhibitors in ODCB solvent, and their partial adsorption on the asphaltene surface are related, and they are a reason of the high value of RMM at $X_I = 0.3$ g/L (see fig.5).

To interpret the interesting phenomena observed in the VPO measurements, we have performed some calculations by means of the canonical MC simulations. In fig.7 the inverse value of the ratio of occupied sites of the model asphaltene surface, $S^{-1}$, is presented. In the part (a) of the figure, we study the influence of the head-solvent phobicity on the inhibitor molecule adsorption. As seen, the increasing of the non-additivity parameter $\Delta_H$ (which corresponds to the decreasing of head-solvent affinity) promotes the inhibitor adsorption on the surface. Such behavior is explained by the entropy gain in the system when the head of the AFI molecule is located on the asphaltene surface liberating the accessible volume in the system. That is why the M2 molecule is more efficient than M3, as shown in fig.2, i.e. the more polar head of M2 has less affinity with toluene if compared to M3 molecule head. Besides, in fig.7a one can see that when $\Delta_H>0$, there is a certain value of $X^I$ after which further augmenting of the inhibitor concentration leads to the increasing of $S^{-1}$ (adsorption lowering); such behavior is also akin to the results presented in fig.2.

In the part b of the fig.7, analysis of different AFI molecules and distances between asphaltene surfaces is presented. Almost for all conditions considered, we have found decreasing of $S^{-1}$ if the concentration of inhibitors augments at the beginning. This means that adsorption (or inhibitor efficiency) increases. However, further increasing of $X^I$ leads to some saturation of $S^{-1}$ value and also to its increasing. Interesting to mention, that the forms of the $S^{-1}$ vs $X^I$ curves tentatively resembles those of the fig.3.

First of all, let us consider in fig. 7b the set of curves that correspond to the model inhibitor molecules with the following parameters: m = 3, $\Delta_H$ = 0.8, and $\Delta_T$ =0. These parameters describe the molecules with the highly lyophobic (phobic with respect to the solvent) head part and neutral tail part. We have considered these molecules at different concentrations when the model asphaltene surfaces are separated by the distances 25 and 40. Almost any effect of the distance has been found. Up to $X^I = 0.04$ the adsorption of the inhibitors on the asphaltene surface increases; however, for higher values of $X^I$ we can observe the slow decreasing of inhibitor adsorption. The snapshots presented in fig.8a clarify this tendency. In the top part of the figure the notable adsorption of the inhibitors is observed, while only few bonds between molecular head and surface active sites are observed in the bottom part of the figure. As seen, the increasing of $X^I$ leads to the inhibitor self-assembly, being the occulting of the phobic head part the mechanism of such phenomena. Thus, adsorption of the inhibitors on the asphaltene surface competes with their self-assembly in the bulk of the solvent [34,35]. Besides, decreasing of the $L_X$ makes possible the formation of AFI bridges between two asphaltenes as shown in Fig.8b, which also can deteriorate the expected AFI efficacy. In our opinion, this can be a main reason why the value of RMM rises when $X_I$ increases in fig.3.

On the other hand, in fig.7b we also study the influence of AFI tail length: Diamonds show the MC results for the AFI molecules with m = 4 ($\Delta_H$ = 0.8, and $\Delta_T$ =0). These molecules manifest lower adsorption (and inhibition efficiency) if compared to AFI's with m = 3. Such tendency is in accord with the VPO measurements presented in fig.3. Namely, the molecule M2 has slightly longer aliphatic tail if compared to NP [22], and NP manifests somewhat better inhibition efficiency than M2 in fig.3.

However, the mechanism just described above does not describe the results of VPO measurements in asphaltene-ODCB-inhibitor system presented in fig.5, where unexpected agglomeration of asphaltene is promoted by the inhibitors that in fact should have prohibited it. To consider theoretically the behavior of the inhibitors in the ODCB solvent one needs to model high phobic tail and less phobic head part of the inhibitor, if compared with the toluene. Therefore, the following set of parameters have been considered: $\Delta_H$ = 0.0, and $\Delta_T$ =0.4. The squares in the fig.7b indicate the relatively low adsorption of such inhibitors on the asphaltene surfaces, high values of $S^{-1}$. The snapshot of the typical MC configuration is presented in fig.9. Unlike the previous case of the less polar solvent, this time we observe the micel-like structures, in which tails try to omit contact with the solvent. More interesting fact is, that such inhibitor complexes have the AFI head groups on their surfaces, and potentially may serve as active "docks" to which asphaltenes can attach. If these complexes are stable, large asphaltene-inhibitor agglomerates can be formed. We suggest, that it was observed in the fig. 5, i.e. both AFI molecules, NP and M2, have self-organized promoting asphaltene accumulation. Two branches of RMM may be attributed to the different size distribution and morphology of the inhibitor complexes which strongly depends on the temperature and concentration fluctuations [10,34].

## Conclusions

Recently, Porte *et al.* [12] has proposed a new general description of the asphaltene agglomeration and precipitation, reconsidering a lot of experimental features previously reported on asphaltene solubilization. According to their approach, in good apolar solvents, aggregation is driven by strong specific forces due to the interaction sites located at the periphery of the asphaltene molecules, and in bad apolar solvents, weak nonspecific dispersion attraction between the aggregates determine precipitation.

Taking into account such description of asphaltene agglomeration, in the presented article we have studied the efficiency of the flocculation inhibitors for the Puerta Ceiba asphaltene solution in toluene and o-dichlorobenzene. Nonylphenol and two new inhibitor molecules (synthesized in our laboratory) have been investigated by means of the vapor pressure osmomentry technique. It has been observed that the asphaltene self-associate into relatively small particles with an aggregation number around 10. Application of the small concentration of inhibitors reduces notably the aggregation number in toluene at 50°C. When inhibitor concentration increases their efficacy diminishes significantly. The same inhibitors promote the formation of asphaltene agglomerates in o-dichlorobenzene at 90°C. Additional VPO measurements of the inhibitor molar mass at its different concentrations, suggest that inhibitor molecules (i) significantly self-associate in the more polar solvent (o-dichlorobenzene), which can be a reason of the asphaltene adsorption (and enhanced agglomeration) on the surface of such inhibitor complexes; (ii) self-associate occulting their head polar part in the less polar solvent (toluene), which may be a reason of the reduced inhibitor adsorption on the asphaltene surface, and as a consequence, the worsening of the inhibition efficacy.

The results of the theoretical modeling of the inhibitor adsorption on the asphaltene surface in different solvents have allowed us to describe the possible mechanisms of the efficient (non-efficient) performance of the asphaltene flocculation inhibitors. In agreement with the VPO experiment, the simulation results show that the inhibitor molecules with more polar head manifest higher adsorption on the asphaltene surface in apolar solvent, while at high concentration they may prefer get-together in the bulk than adsorption on the surface.

In our work we have studied the inhibition mechanism in aliphatic solvents, which simulate the crude oils apolar medium, and may not be completely representative of crude oils. Once operation variables in the field may be very distinct from laboratory experiments, the necessity of studies that establish a relationship between the additives effectiveness and their capacity in the field, is obvious. This problem is now under study in our laboratory.

**Acknowledgment**. The authors thank the financial support provided by *Programa de Ingeniería Molecular* (IMP, México) under Projects No. D.31519/D.00405/D.00406

# Figure Captions

Fig. 1. Schematic representation of the computational cell. The length of the cell along the Y direction is $L_y = 20 \times r_{ss} = 30$, were $r_{ss}$ is a distance between the asphaltene surface sites (small black circles).

Fig.2 VPO results of the molar mass of Puerto Ceiba asphaltenes with different amount of inhibitor M2 (part a) and M3(part b) in toluene at 50°C.

Fig. 3 VPO results of the reduced molar mass, RMM, of Puerto Ceiba asphaltenes with different amount of inhibitor NP (part a ) and M2 (part b) in toluene at 50°C. Circles and squares depict asphaltene concentration $C_A$=6 g/L and 15 g/L, respectively.

Fig. 4 VPO results of the inhibitor molar mass in toluene at 50°C: NP (circles ) and M2 (stars).

Fig. 5 VPO results of the reduced molar mass, RMM, of Puerto Ceiba asphaltenes with different amount of inhibitor NP (circles) and M2 (stars) in ODCB at 90°C; asphaltene concentration is $C_A$= 6 g/L .

Fig. 6 VPO results of the inhibitor molar mass in ODCB at 90°C: NP (circles) and M2 (stars).

Fig. 7 Simulation results of the inverse ratio of occupied surface sites, 1/S, as a function of the model inhibitor concentration, $X^I$. Part (a) corresponds to $L_X$=25, m=3, $\Delta_{TT}$ =0.4, $\Delta_T$ =0.4, and $\Delta_H$ = 0.0 (triangles), 0.4 (squares), and 0.8 (diamonds). Part (b) corresponds to $\Delta_{TT}$ =0.8, m=4 (diamonds), m=3 (other symbols); $\Delta_H$ = 0.8, and $\Delta_T$ =0 (white and gray symbols); $\Delta_H$ = 0, $\Delta_T$ =0.4, and $L_X$=30 (black squares); $L_X$=25 (stars) and , $L_X$=40 (diamonds and circles).

Fig. 8 (a) Snapshots of typical configurations of the model system at two different inhibitor concentrations, $X^I$ = 0.031 (top part) and 0.05 (bottom part); $L_X$=40, m=3, $\Delta_H$ = 0.8, and $\Delta_T$ =0. (b) Snapshot of typical configuration of the model system at $X^I$ = 0.056; $L_X$=18, m=3, $\Delta_H$ = 0.8, and $\Delta_T$ =0.

Fig. 9 Snapshot of typical configuration of the model system at $X^I$ = 0.06, $L_X$ =30, m=3, and $\Delta_H$ = 0, $\Delta_T$ =0.4.

# Table and Figures

**Table 1.** SARA Analysis and °API of Puerto Ceiba crude oil.

| Saturates (%wt) | Aromatics (%wt) | Resins (%wt) | Asphaltenes (%wt) | °API |
|---|---|---|---|---|
| 41.7 | 34.2 | 21.8 | 2.3 | 36 |

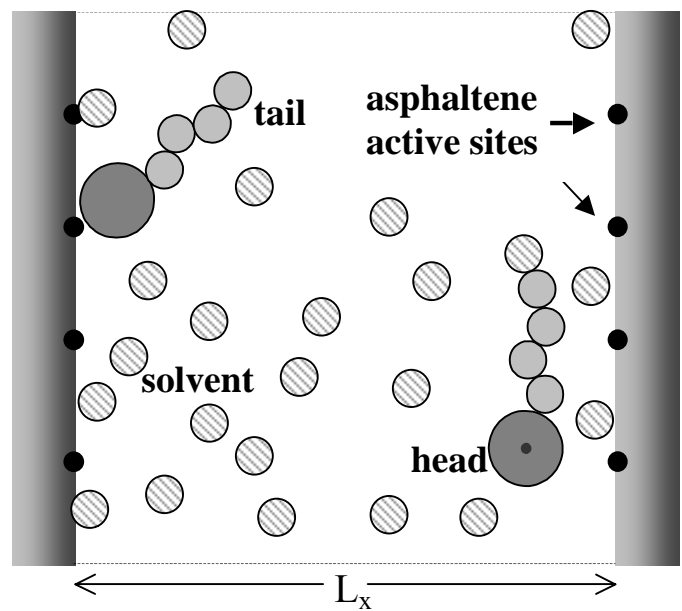

**Figure 1**

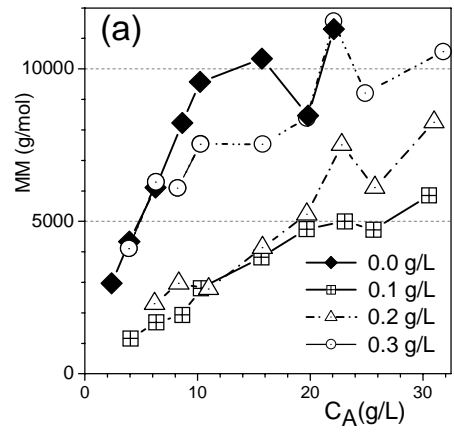

**Figure 2a**

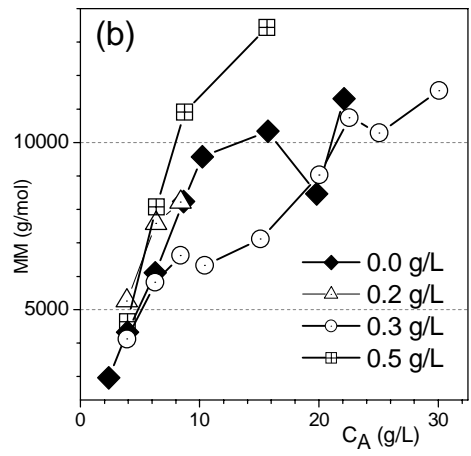

**Figure 2b**

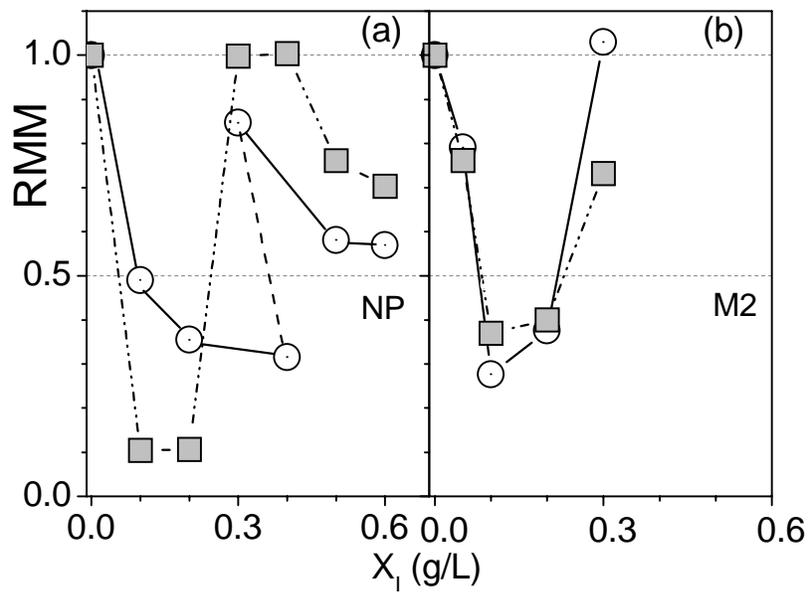

**Figure 3**

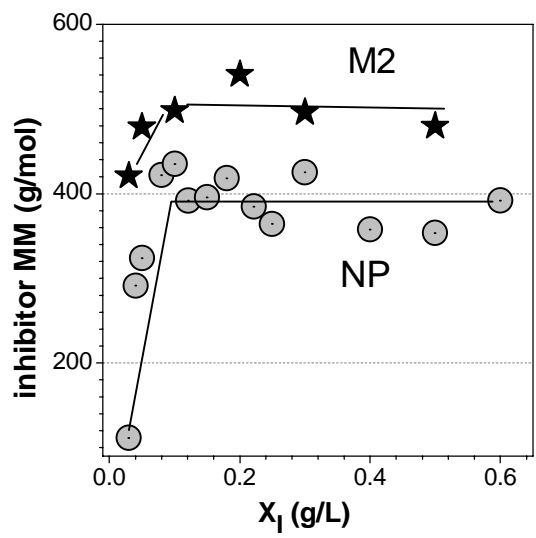

**Figure 4**

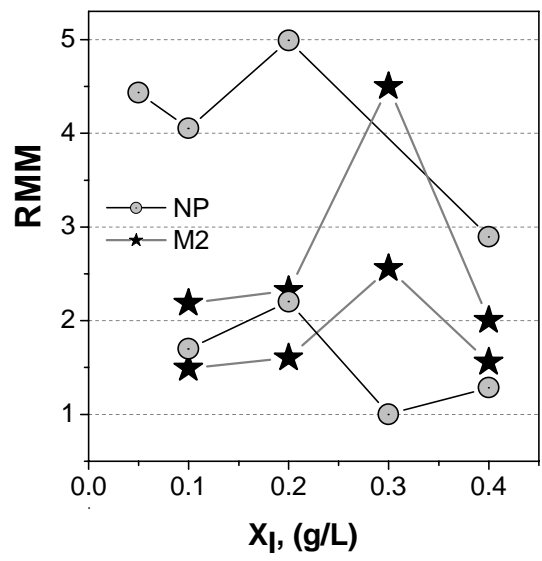

**Figure 5**

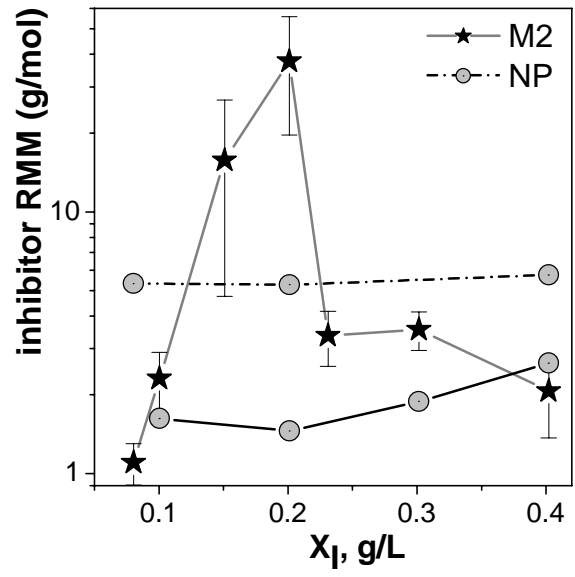

Figure 6

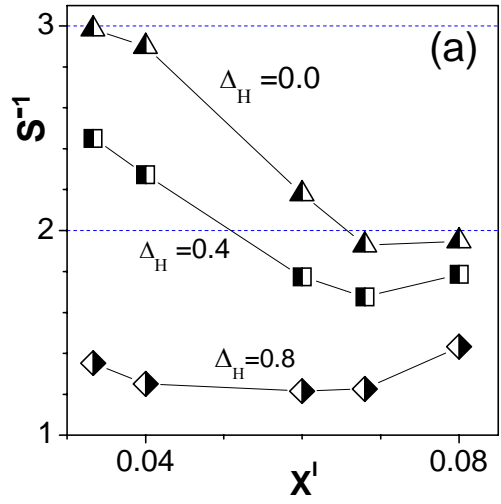

**Figure 7a**

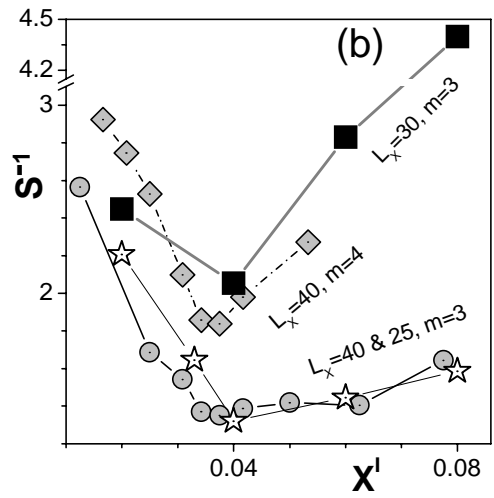

**Figure 7b**

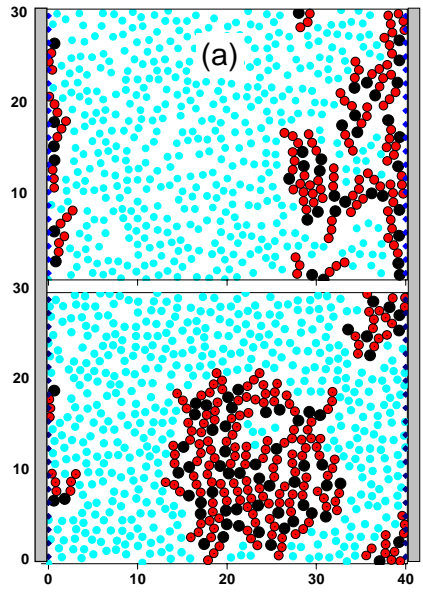

**Figure 8a**

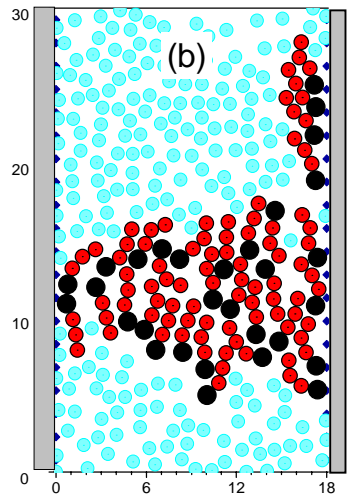

**Figure 8b**

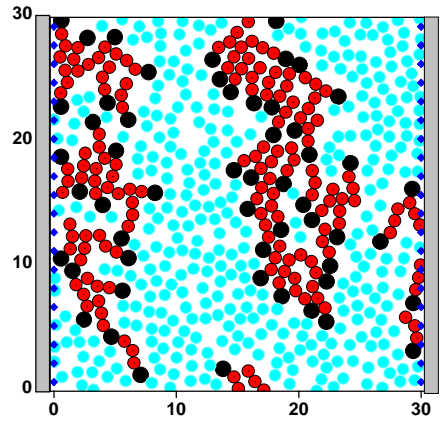

**Figure 9**